\documentclass{aastex62}

\graphicspath{{./}}

\received{XXX}
\revised{XXX}
\accepted{XXX}

\submitjournal{ApJ}

\usepackage{rotating}

\shorttitle{Complexity of magnetic exhausts}
\shortauthors{Miranda et al.}

\begin{document}

\title{Complexity of magnetic-field turbulence at reconnection exhausts in the solar wind at 1 AU}

\correspondingauthor{Rodrigo A. Miranda}
\email{rmiracer@unb.br}

\author[0000-0002-9861-0557]{Rodrigo A. Miranda}
\affil{UnB-Gama Campus and Institute of Physics, University of Bras\'ilia,
Bras\'ilia, DF 70910-900, Brazil}

\author[0000-0003-3381-9904]{Juan A. Valdivia}
\affil{Departamento de F\'isica, Facultad de Ciencias, Universidad de Chile, Santiago, Chile}
%\affil{Department of Physics, Faculty of Sciences, University of Chile, Santiago, Chile}

\author[0000-0002-8932-0793]{Abraham C.-L. Chian}
\affiliation{School of Mathematical Sciences, University of Adelaide, Adelaide, SA 5005, Australia}
\affiliation{National Institute for Space Research (INPE), S\~ao Jos\'e dos Campos, SP 12227-010, Brazil}

\author[0000-0002-3435-6422]{Pablo R. Mu\~noz}
\affiliation{Department of Physics and Astronomy, University of La Serena, La Serena, Chile}

\begin{abstract}

 Magnetic reconnection is a complex mechanism that converts magnetic energy into particle kinetic energy and plasma thermal energy in space and astrophysical plasmas. In addition, magnetic reconnection and turbulence appear to be intimately related in plasmas. We analyze the magnetic-field turbulence at the exhaust of four reconnection events detected in the solar wind using the Jensen-Shannon complexity-entropy index. The interplanetary magnetic field is decomposed into the LMN coordinates using the hybrid minimum variance technique. The first event is characterized by an extended exhaust period that allows us to obtain the scaling exponents of higher-order structure functions of magnetic-field  fluctuations. By computing the complexity-entropy index we demonstrate that a higher degree of intermittency is related to lower entropy and higher complexity in the inertial subrange. We also compute the complexity-entropy index of three other reconnection exhaust events. For all four events, the $B_L$ component of the magnetic field displays a lower degree of entropy and higher degree of complexity than the $B_M$ and $B_N$ components. Our results show that coherent structures can be responsible for decreasing entropy and increasing complexity within reconnection exhausts in magnetic-field turbulence.

\end{abstract}


\section{Introduction} \label{sec:intro}

  Magnetic reconnection in plasmas refers to the process in which magnetic energy is converted to particle kinetic and thermal energy, resulting in a change of topology of the magnetic-field lines \citep{yamada_etal:2010, treumann_baumjohann:2013, lazarian_etal:2015}. The study of magnetic reconnection is key to understand the dynamics of solar flares, coronal mass ejections, rope-rope magnetic reconnection in the solar wind, and the interaction between solar wind and planetary magnetospheres. In addition, magnetic reconnection, turbulence, and intermittency seem to be intrinsically related in plasmas in a complex manner, hence, they need to be studied in relation to each other.

  The solar wind is a natural laboratory for the study of magnetic reconnection. The conversion of magnetic energy into particle kinetic energy during the reconnection process leads to the formation of magnetic exhausts. The properties of magnetic exhausts have been studied recently using observational data. For example, \cite{enzl_etal:2014} performed a statistical survey of 418 reconnection exhausts detected by the Wind spacecraft. They showed that the magnetic flux available for reconnection and the reconnection efficiency increase with the magnetic shear angle. \cite{mistry_etal:2015} analyzed data from different spacecraft sampling oppositely directed reconnection exhausts of three different reconnection events. They showed that bifurcated current sheets are clearly observed when the spacecraft is located at a distance greater than $\sim 1000 d_i$ from the 
  X-line, where $d_i$ is the ion skin depth. \cite{chian_etal:2016} demonstrated that magnetic reconnection at the interface of two magnetic flux ropes provides an origin of intermittent magnetic-field turbulence in the solar wind. The statistics of 188 reconnection exhausts was studied by \cite{mistry_etal:2017}. They showed that the guide magnetic field within the exhaust is enhanced, and the plasma density and ion temperature at the exhaust increase as a function of the inflow plasma beta and the guide field. Numerical simulations have also been used to understand the properties of the turbulent plasma in reconnection exhausts. \cite{pucci_etal:2017} performed three-dimensional (3D) particle-in-cell (PIC) simulations of magnetic reconnection, and showed that the turbulence at the outflows is anisotropic, and that the energy exchange and dissipation is concentrated at the interface between the ejected plasma and the ambient plasma. The outflow region has also been identified by \cite{lapenta_etal:2018} as a source of instabilities that feeds a turbulent cascade and secondary reconnection sites in 3D PIC simulations of reconnection with a weak guide field. \cite{adhikari_etal:2020} computed the scaling laws of energy spectra and second-order structure functions of 2.5D PIC simulations. They demonstrated that the inflow region displays a lower degree of turbulence compared to the diffusion, exhaust, separatrix, and island regions resulting from the reconnection process. Hence, there is a need to quantitatively characterize the turbulent behavior in these regions; for example, using a complex system approach; which can complement theoretical and simulation efforts.

  In this respect, the Jensen-Shannon (J-S) complexity-entropy index is a statistical tool that allows to distinguish noise from chaos~\citep{rosso_etal:2007}. It has been successfully applied to data from experiments with electronic oscillators~\citep{soriano_etal:2011}, stock market data~\citep{zunino_etal:2009}, the Southern Oscillation index~\citep{bandt:2005}, and heart rate variability \citep{bian_etal:2012}, among others. For a detailed list of applications see \cite{riedl_etal:2013}. \cite{weck_etal:2015} computed the J-S index of the interplanetary magnetic-field data detected by the Wind spacecraft, magnetic-field data of the Swarthmore Spheromak Experiment (SSX), and the ion saturation current data at the edge of the Large Plasma Device (LAPD). They showed that the Wind data displays high-entropy and low-complexity similar to stochastic signals; whereas the SSX and the LAPD data display intermediate entropy and high complexity due to the lower number of degrees of freedom of the experimental devices, and the confined nature of the experiments compared to the interplanetary magnetic-field data. The J-S index was also applied to solar wind data collected by the Helios, Wind, and Ulysses spacecraft by \cite{weygand_kivelson:2019}. Several intervals were selected, including slow and fast solar wind, interplanetary coronal mass ejections, and corotating interactions regions. They also obtained J-S index values characteristic of stochastic fluctuations, and showed that the complexity decreases and the entropy increases with the distance from the Sun.

  In this paper we characterize the complexity-entropy of magnetic-field data of four reconnection exhausts detected in the solar wind. For the first event, we show that intermittency and multifractality are related to the degree of entropy and complexity. By projecting the magnetic field into the LMN coordinates \citep{sonnerup_calil:1967, gosling_phan:2013}, we show that the L component displays lower entropy and higher complexity than the M and N components. Our paper is organized as follows. Section \ref{sec_events} describes briefly the four magnetic reconnection events. Section \ref{sec_methods} presents the methods employed for the data analysis. The results are presented in Section \ref{sec_results}, and a discussion and conclusions are given in Section \ref{sec_discussions}. 

\section{Magnetic reconnection events} \label{sec_events}

We analyze four reconnection exhausts detected at 1 AU. The timing of each interval is indicated in Table \ref{table1}. Events 1 and 3 are magnetic reconnection exhausts detected by Wind at the interior of a magnetic cloud associated with an ICME, with a main shock arrival observed at 1:13 UT on 30 December 1997 and at 8:55 UT on 22 November 1997, respectively. Event 2 is a magnetic reconnection exhaust detected by Wind after the passage of an ICME with a main shock arrival observed at 7:18 UT on 12 November 1998. Event 4 is a reconnection exhaust detected by Cluster on 2 February 2002 \citep{phan_etal:2006}. This exhaust is the result of the magnetic reconnection between a small-scale interplanetary magnetic flux rope (IMFR) and an intermediate-scale IMFR \citep{chian_etal:2016}. We use the magnetic-field data from Wind at a resolution of 11 Hz for events 1, 2 and 3, whereas for event 4 we employ the data from Cluster at a resolution of 22 Hz.

The timing of events 1, 2, and 3 is based on the supplement table of 188 magnetic reconnection exhausts studied by \citet{mistry_etal:2017}. These events are the longest exhaust intervals during which the IMF experiment onboard Wind was operating at 11 Hz, resulting in a sufficient number of data points for analysis. For event 4, we use the data from four Cluster spacecraft and apply the curlometer technique to compute the modulus of the current density $\mathbf{J}$. Since the exhaust is bounded by a bifurcated current sheet in the Petschek model of magnetic reconnection \citep{gosling_szabo:2008}, the exhaust interval of event 4 is defined using the timing of the main peaks of $|\mathbf{J}|$. Table \ref{table1} also indicates the number of data points available from each interval.

\section{Methods} \label{sec_methods}

The vector magnetic field is projected on the LMN coordinate system by applying the hybrid minimum variance analysis (MVA) \citep{gosling_phan:2013, mistry_etal:2015, mistry_etal:2017, hietala_etal:2018}. The $L$ component is given by the direction of maximum variance and is related to the exhaust outflow direction, the $M$ direction is related to the reconnection guide field direction, and the $N$ component is the direction of minimum variance related to the normal of the current sheet. The $N$ direction can be obtained as

\begin{displaymath}
  \mathbf{\hat{e}}_N = \frac{\mathbf{B}_1 \times \mathbf{B}_2}{|\mathbf{B}_1 \times \mathbf{B}_2|},
\end{displaymath}

\noindent where $\mathbf{B}_1$ and $\mathbf{B}_2$ are the magnetic-field vectors immediately adjacent to the exhaust boundaries. The $M$ direction is given by

\begin{equation} \label{eq_em_MVA}
  \mathbf{\hat{e}}_M = \mathbf{\hat{e}}_N \times \mathbf{\hat{e}}_{L'},
\end{equation}

\noindent where $\mathbf{\hat{e}}_{L'}$ is the maximum variance direction obtained from the classical MVA \citep{sonnerup_calil:1967}; and the $L$ direction is

\begin{equation} \label{eq_el_MVA}
  \mathbf{\hat{e}}_L = \mathbf{\hat{e}}_M \times \mathbf{\hat{e}}_N.
\end{equation}

\noindent Note that, from Eq. (\ref{eq_em_MVA}), $\mathbf{\hat{e}}_N$ and $\mathbf{\hat{e}}_{L'}$ are not necessarily orthogonal, whereas $\mathbf{\hat{e}}_N$ and $\mathbf{\hat{e}}_M$ are orthogonal. From Eq. (\ref{eq_el_MVA}), $\mathbf{\hat{e}}_L$ is made orthogonal to $\mathbf{\hat{e}}_N$ and $\mathbf{\hat{e}}_M$. The hybrid MVA is more reliable than the MVA \citep{knetter_etal:2004} because the MVA can fail to separate the intermediate and minimum variance directions properly, giving unrealistic values for the $N$ component \citep{mistry_etal:2017}.

We compute the power spectral density (PSD) of the $B_L$, $B_M$, and $B_N$ components using the Welch method \citep{welch:1967} which allows us to reduce the error of the spectrum estimate. The compensated PSD is obtained by multiplying the original PSD by $f^{+5/3}$. The inertial subrange can be identified as a frequency range in which the compensated PSD is nearly horizontal.

The large number of data points within the exhaust interval of event 1 allows us to compute higher $p$th-order structure functions $S_p(\tau) = \left< | B_i(t + \tau) - B_i(t) |^p \right>$, for $i = L, M, N$~\citep{deWit:2004, miranda_etal:2013}. The scaling exponents of structure functions can be computed from $S_p(\tau) \sim \tau^{\alpha(p)}$, to quantify the departure from self-similarity (i.e., multifractality). The scaling exponents $\alpha(p)$ can be obtained by plotting $S_p(\tau)$ as a function of $\tau$ in log-log scale, and applying a linear fit within the inertial subrange. The inertial subrange is identified as the range of scales in which $\alpha(p = 3) = 1$. However, the number of scales within the inertial subrange can be small and difficult to determine, especially for short time series. Therefore, the numerical values of $\alpha(p)$ will be affected by a large statistical error. For this reason we apply the Extended Self-Similarity technique \citep{benzi_etal:1993} by assuming $S_p(\tau) \sim [S_3(\tau)]^{\zeta(p)}$, where $\zeta(p) \sim \alpha(p)/\alpha(3)$. The value of $\zeta(p)$ can be estimated by plotting $S_p(\tau)$ as a function of $S_3(p)$ in log-log scale. The ESS technique results in a larger range of scales in which $\zeta(p = 3) = 1$, thus reducing the statistical error and producing a more robust value for the computed scaling exponents.

We compute the Jensen-Shannon complexity index, in which a probability distribution function (PDF) of ordinal patterns is obtained from the magnetic-field data within the exhaust interval. This PDF represents the frequencies of occurrence of all possible ordinal patterns of length $d$ \citep{bandt_pompe:2002, weck_etal:2015}. For example, suppose that the time series of a component of the magnetic field starts with \{-2.67, 10.80, 1.72, -2.40, 11.21, ...\}. The first $d$-tuple of length $d = 3$ is (-2.67, 10.80, 1.72) and the corresponding ordinal pattern, in ascending order, is (1, 3, 2) because $-2.67 < 1.72 < 10.80$. The second 3-tuple is (10.80, 1.72, -2.40) and the ordinal pattern is (3, 2, 1). For a time series of length $K$, there are $K - d + 1$ $d$-tuples and $d!$ possible permutations of a $d$-tuple. The PDF of ordinal patterns is obtained by counting the number of occurrences of each possible permutation of ordinal patterns within the time series

\begin{displaymath}
  p_i = \frac{\#\{s |s  \textrm{ is the i$^{th}$ ordinal pattern}\}}{K - d + 1},
\end{displaymath}

\noindent where ``$\#$'' stands for ``number'', and $K - d + 1 > d!$. The Shannon entropy is given by

\begin{equation} \label{eq_shannon_entropy}
  S(P) = - \sum_{i = 1}^{d!} p_i \ln(p_i),
\end{equation}

\noindent where $P$ represents the PDF of ordinal patterns. The Shannon entropy is equal to zero if, for any $i$, $p_i = 1$ and $p_j = 0, j \ne i$. This case represents a completely ordered system. Conversely, the Shannon entropy will be maximum if all possible ordinal patterns have the same probability. In this case $S(P_e) = \ln(d!)$, where $P_e$ represents the uniform distribution. The normalized Shannon entropy can be written as

\begin{equation} \label{eq_normalized_shannon}
  H(P) = \frac{S(P)}{S(P_e)} = \frac{S(P)}{\ln(d!)}.
\end{equation}

\noindent Similarly, the Jensen's divergence measures the ``disequilibrium'', or the ``distance'' between a distribution $P$ and the uniform distribution $P_e$ \citep{martin_etal:2006, rosso_etal:2007}

\begin{equation} \label{eq_jensen_divergence}
  Q_J(P, P_e) = Q_0 \left[ S \left( \frac{P + P_e}{2} \right) - \frac{1}{2} S(P) - \frac{1}{2} S(P_e) \right],
\end{equation}

\noindent where $Q_0$ is a normalization constant given by \citep{martin_etal:2006}

\begin{displaymath}
  Q_0 = - \frac{1}{2} \left[ \frac{K+1}{K} \ln(K + 1) - 2 \ln(2K) + \ln(K) \right].
\end{displaymath}

\noindent The Jensen-Shannon complexity is then given by

\begin{equation} \label{eq_jensen_shannon_complexity}
  C_J^S = Q_J(P, P_e) H(P).
\end{equation}

\noindent The pair $(H, C_J^S)$ can be represented in a plane called the complexity-entropy (C-H) plane. This plane can be separated into three regions, namely, a low-entropy and low-complexity region corresponding to highly predictable systems, an intermediate-entropy and high-complexity region corresponding to unpredictable systems with a large degree of structure, and a high-entropy and low-complexity region corresponding to stochastic-like processes \citep{rosso_etal:2007}.

The number of data points $K$ needed to compute Eqs. (\ref{eq_normalized_shannon}) and (\ref{eq_jensen_shannon_complexity}) reliably is \citep{amigo_etal:2008,riedl_etal:2013}

\begin{equation} \label{eq_number_datapoints_permutation}
  K > 5 d!.
\end{equation}

\noindent A large value of the $d$ parameter can result in unreliable statistics, whereas a small value of $d$ can result in an overestimated value of Eq. (\ref{eq_jensen_shannon_complexity}) \citep{gekelman_etal:2014}. We set $d$ to the maximum value for which Eq. (\ref{eq_number_datapoints_permutation}) is satisfied for all intervals, as recommended by \cite{amigo_etal:2008} and \cite{riedl_etal:2013}. For $d = 5$ we need $K >$ 600, whereas for $d = 6$, $K > 3600$. From Table \ref{table1}, all intervals have enough data points to satisfy Eq. (\ref{eq_number_datapoints_permutation}) with $d = 5$.

\section{Intermittency and complexity in reconnection exhausts} \label{sec_results}

\begin{figure}
  \begin{center}
    \includegraphics[height=0.9\textheight]{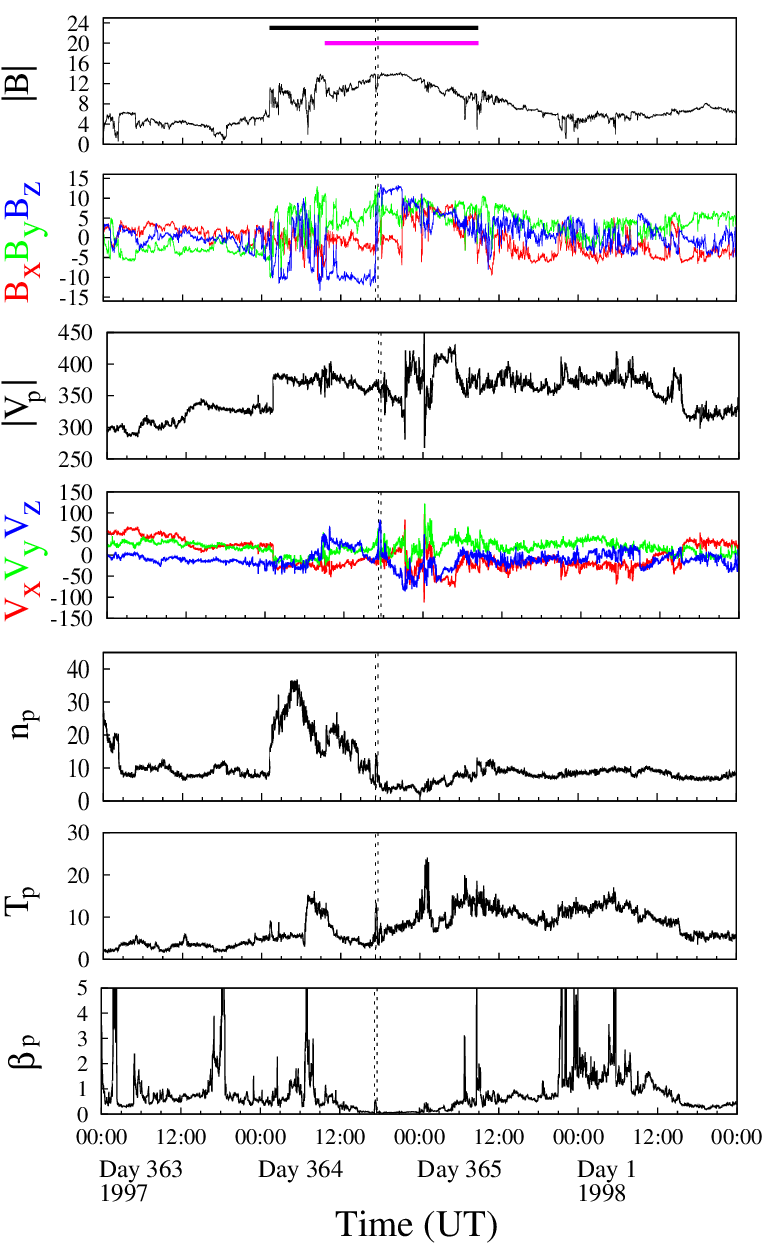}
  \end{center}
  \caption{Interplanetary magnetic field and plasma data detected by the Wind spacecraft during the passage of an ICME on 30 December 1997 (Julian day 364). From top to bottom: the modulus of the magnetic field $|\mathbf{B}|$ (nT), the three components of the magnetic field $B_x$, $B_y$, and $B_z$ in GSE coordinates (nT), the modulus of the proton velocity $|\mathbf{V}_p|$ (km/s), the three components of the proton velocity $V_x$, $V_y$, and $V_z$ in GSE coordinates (km/s), the proton density $n_p$ (cm$^{-3}$), the proton temperature $T_p$ (eV) and the proton beta $\beta_p$. The horizontal black line in the top panel indicates the ICME interval, and the violet horizontal line indicates the magnetic cloud interval. The vertical dashed lines indicate the reconnection exhaust interval.}
  \label{fig1}
\end{figure}

We start our analysis by describing the solar wind conditions around event 1. Figure \ref{fig1} shows an overview of the plasma parameters observed by the MFI an SWE instruments onboard Wind, namely, the modulus of the magnetic field $|\mathbf{B}|$ (nT); the three components of the magnetic field $B_x$, $B_y$, and $B_z$ (nT) in the GSE coordinates; the modulus of the proton velocity $|\mathbf{V}_p|$ (km/s); the proton density $n_p$ (cm$^{-3}$); the proton temperature $T_p$ (eV); and the proton beta $\beta_p = 8 \pi n_p K_B T_p / |\mathbf{B}|^2$, where $K_B$ is the Boltzmann constant. This event is characterized by a main shock arrival detected at 1:13 UT on 30 December 1997, and a magnetic cloud from 9:35 UT on 30 December to 8:51 UT on 31 December \citep{nieveschinchilla_etal:2018}. This magnetic cloud is characterized by an increase on $|\mathbf{B}|$, a rotation of the magnetic-field direction, and a decrease in $T_p$ and $\beta_p$. The horizontal lines indicate the boundaries of the ICME (black) and the magnetic cloud (violet), and the vertical dashed lines indicate the reconnection exhaust interval. Note that the value of $\beta_p$ is low in Fig. \ref{fig1} outside the reconnection exhaust because this event occurs at the interior of a magnetic cloud.

Figure \ref{fig2} shows a detailed view of the plasma parameters around event 1. The magnetic reconnection event is characterized by a decrease of $|\mathbf{B}|$, a change of the $B_z$ magnetic-field component in the GSE coordinates, and an increase of the proton beta $\beta_p$. The magnetic-field components in the LMN coordinates are also shown. The reconnection event is also characterized by the corresponding increases of the $V_z$ velocity component (in GSE), $n_p$, and $T_p$. The reconnection exhaust interval, bounded by the two vertical dashed lines, has a duration of 1343 seconds which gives 14783 data points within the exhaust (as indicated in Table \ref{table1}).

The time series of the $B_L$ component shown in Fig. \ref{fig2} displays discontinuous ``jumps'' near the boundaries of the reconnection exhaust interval, associated with the magnetic-field reversal that is required for the magnetic reconnection to occur. Since the magnetic shear occurs at a scale larger than the selected interval, we apply a trend removal technique to the time series of $B_L$. This is achieved by computing a third-order polynomial fit to the time series of $B_L$, and removing the resulting fit from the original time series. Figure \ref{fig3}(a) shows the time series of $B_L$ after detrending, represented by $B_L^*$. We have also applied the same procedure to obtain the time series of the $B_M^*$ and $B_N^*$ components, shown in Figs. \ref{fig3}(b) and \ref{fig3}(c), respectively. Hereafter, we will analyze the time series of $B_L^*$, $B_M^*$, and $B_N^*$, and refer them simply as $B_L$, $B_M$, and $B_N$, respectively.

Figure \ref{fig4}(a) shows the power spectral density (PSD) of the $B_L$, $B_M$, and $B_N$ components of the magnetic field. The dashed line in the upper panel indicates the $-5/3$ power-law scaling within the inertial subrange. This interval was obtained by plotting the compensate PSD shown in the lower panel of Fig. \ref{fig4}(b). In this figure the range of scales in which the PSDs of $B_L$, $B_M$, and $B_N$ are nearly horizontal is indicated by vertical dashed lines.

We compute the scaling exponents $\zeta$ of the structure functions with the ESS technique (i.e., $S_p(\tau) \sim [S_3(\tau)]^{\zeta(p)}$), within the inertial subrange identified by the compensated PSD. Figure \ref{fig5} shows the structure functions before and after applying the ESS technique to the time series of the $B_L$ component. The inertial subrange in Fig. \ref{fig5}(a), shown as a grey background, is obtained from the compensated PSD of Fig. \ref{fig4}(b), and coincides with the interval in which $\alpha(p = 3) = 1$. The grey background of Fig. \ref{fig5}(b) indicates the extended interval in which the scaling exponents $\zeta$ are obtained. The horizontal black line represents the original inertial subrange. Note that the extended interval cannot include kinetic scales, which starts near the ion cyclotron frequency. From Fig. \ref{fig4}, a spectral break marking the end of the inertial subrange occurs near $f = 0.5$ Hz, which corresponds to a scale of $\tau \sim 2$ s. This scale is outside the shaded region of Fig. \ref{fig5}(a), and corresponds to $S_3(\tau)/S_3(T) = 238$ in the horizontal axis of Fig.  \ref{fig5}(b), which is also outside the interval used to compute the scaling exponents. We have also applied the ESS technique to the structure functions of the $B_M$ and $B_N$ components using the same procedure.

Figure \ref{fig6} shows $\zeta$ as a function of the $p$-th order structure function for $B_L$, $B_M$, and $B_N$. The vertical bars indicate the error in the fit. Intermittency and multifractality within the inertial subrange are responsible for deviations of $\zeta$ from the linear scaling of Kolmogorov's 1941 (hereafter K41) self-similar model. This figure shows clearly that, for higher-order statistics, the $B_L$ component displays a stronger departure from the K41 scaling than $B_M$, which in turn displays a greater departure than $B_N$. Therefore, the $B_L$ component displays a higher degree of multifractality and intermittency than the $B_M$ and the $B_N$ components.

Next, we show how the intermittency of the magnetic-field fluctuations is related to entropy and complexity for event 1. We compute the J-S index (i.e., Eqs. (\ref{eq_normalized_shannon}) and (\ref{eq_jensen_shannon_complexity})) for $B_L$, $B_M$, and $B_N$. Figure \ref{fig7} shows the $d = 5$ C-H plane. The crescent-shaped curves indicate the minimum and maximum values of $C_J^S$ for a given value of $H$. Symbols indicate the ($H$, $C_J^S$) values of three chaotic maps, namely, the logistic map $x_{n+1} = rx_n(1 - x_n)$ with $r = 4$; the skew tent map
\begin{eqnarray*}
  x/w, & \qquad & x \in [0, w[,\\
  (1 - x)/(1 - w), & \qquad & x \in [w, 1],
\end{eqnarray*}
\noindent with $w = 0.1847$; and the H\'enon map
\begin{eqnarray*}
  x_{n + 1} & = & 1 - ax_n^2 + y_n, \\
  y_{n + 1} & = & b x_n,
\end{eqnarray*}
\noindent with $a = 1.4$ and $b = 0.3$. The parameter values of the chaotic maps are the same from \cite{rosso_etal:2007} and \cite{weck_etal:2015}. Their locations on the C-H plane identify the region corresponding to deterministic chaotic behavior. The dotted curve represents the ($H$, $C_J^S$) values of stochastic fractional Brownian motion (fBm). This curve was computed by generating time series of fBm with a Hurst exponent varying within the interval $[0.025, 0.925]$ \citep{maggs_morales:2013}. Smaller Hurst exponents display larger $H$ and lower $C_J^S$ values. The chaotic maps and stochastic signals allow us to illustrate the different regions of the C-H plane.

The ordinal PDFs of $B_L$, $B_M$, and $B_N$ are computed by setting the size of the ordinal pattern to $d$ = 5. We also define an embedding delay $T$, which means that $d$-tuples are sampled on a larger time scale instead of consecutive points. This allows us to relate the computed ($H$, $C_J^S$) values to a given time scale. We set the embedding delay $T=110$ data points. For event 1, this value of $T$ corresponds to a time scale of 10~s, which is within the inertial subrange (see Fig. \ref{fig4}). Similar results were obtained for embedding delay values corresponding to time scales $\in [5, 15]$~s. Figure \ref{fig7} shows that the three magnetic-field components display ($H$, $C_J^S$) values close to the bottom-right region of the C-H plane, which correspond to stochastic-like processes. However, the $B_L$ component displays a lower degree of entropy and a higher degree of complexity than the $B_M$ component, which in turn displays lower entropy and higher complexity than the $B_N$ component. This pattern is also observed when we choose different values of $d$. Setting $d = 4$, the $(H, C_J^S)$ values of the three magnetic-field components are slightly shifted closer to the bottom-right region of the entropy-complexity plane, while their relative positions in the C-H plane with respect to each other are maintained. For $d = 6$, the three values are slightly shifted away from this region, also keeping their relative positions in the C-H plane, demonstrating the robustness of this result. Note that the number of data points of event 1 still satisfy Eq. (\ref{eq_number_datapoints_permutation}) with $d = 6$. The numerical values of ($H$, $C_J^S$) for $d$ = 4, 5, and 6 are given in Table \ref{table2}.

We apply the same analysis to events 2, 3, and 4, except for the computation of scaling exponents, because the number of data points for each of these events is insufficient to guarantee the convergence of high-order statistics. However, as stated in Section \ref{sec_methods}, the number of data points of these events satisfy Eq. (\ref{eq_number_datapoints_permutation}) for $d$ = 5, allowing us to compute $H$ and $C_J^S$. Figure \ref{fig7} shows the entropy-complexity plane of the four reconnection exhausts. For all the analyzed events, the $B_L$ component displays lower values of $H$ and higher values of $C_J^S$, as compared with the $B_M$ and $B_N$ components.

\section{Discussion and conclusions} \label{sec_discussions}

The results shown in Figs. \ref{fig6} and \ref{fig7} suggest that a higher degree of intermittency is related to a decrease of entropy and an increase of complexity. Intermittency is related to the presence of coherent structures in turbulent fluids and plasmas, resulting in non-Gaussian PDFs \citep{sorrisovalvo_etal:2001,chian_miranda:2009}, finite degree of amplitude-phase synchronization \citep{koga_etal:2007, chian_miranda:2009}, and multifractal scaling exponents \citep{bershadskii_sreenivasan:2004, bruno_etal:2007, miranda_etal:2013}. Coherent structures are also responsible for lower values of the Fourier power spectral entropy in 3D compressible MHD simulations of an intermittent dynamo \citep{rempel_etal:2009} and lower values of the spectral power and phase entropies in 3D incompressible MHD simulations of a Keplerian shear flow \citep{miranda_etal:2015}. We have shown that the $B_L$, $B_M$, and $B_N$ components have $H$ and $C_J^S$ values similar to stochastic fluctuations, in agreement with previous analyses of interplanetary magnetic-field data \citep{weck_etal:2015, weygand_kivelson:2019}. Our results indicate that, within magnetic reconnection exhausts, coherent structures are responsible for decreasing entropy and increasing complexity. We note that the $H$ and $C_J^S$ values of the $B_L$, $B_M$, and $B_N$ components are consistent with those of fBm with Hurst exponents 0.525, 0.425, and 0.335, respectively. These values can provide additional quantitative constraints to simulation and modeling of magnetic reconnection in plasmas.

By comparing the J-S index of the $B_L$, $B_M$, and the $B_N$ magnetic-field components we have shown that, for the four events selected, the $B_L$ component has lower entropy and higher complexity than the $B_M$ component; and the $B_M$ component has lower entropy and higher complexity than the $B_N$ component. From the analysis of event 1 it follows that the $B_L$ component is more intermittent than $B_M$ and $B_N$ due to coherent structures within the inertial subrange. Coherent structures and intermittency are also related to energy dissipation in anisotropic MHD turbulence \citep{muller_etal:2003, miranda_etal:2013}. Our results indicate that the energy dissipation in these four magnetic reconnection exhaust events is strongest in the $B_L$ component. For events 2, 3, and 4, high-order statistics and scaling exponents cannot be computed reliably due to the small number of data points. High-resolution data from instruments operating on higher cadence modes would be needed for the convergence of higher-order statistics. However, we have shown that similar results can be obtained by computing the J-S index. Since magnetic exhausts in the solar wind at 1 AU usually have a short duration \citep{enzl_etal:2014}, the J-S index can be a useful tool for the analysis of the magnetic-field turbulence within exhausts.

\begin{figure}
  \begin{center}
    \includegraphics[height=0.9\textheight]{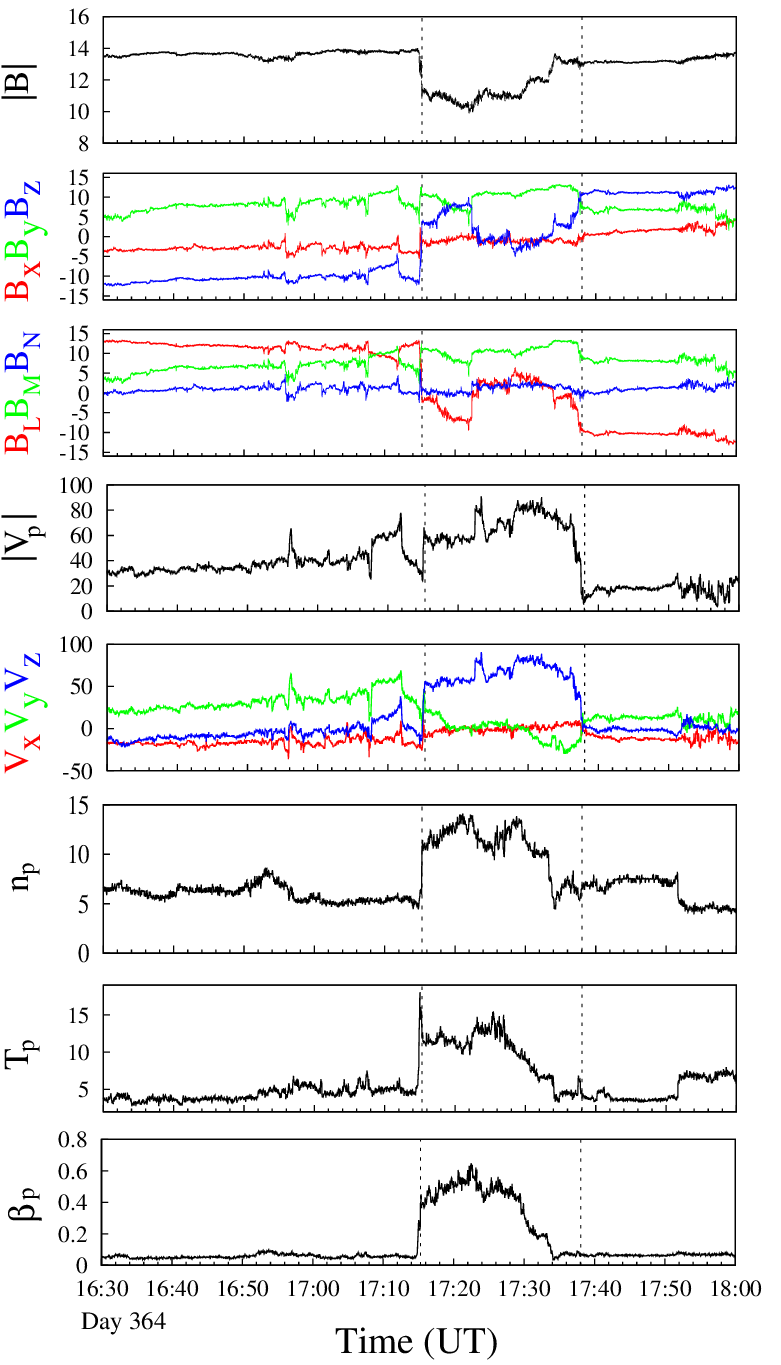}
  \end{center}
  \caption{Magnetic reconnection exhaust (event 1) detected on 30 December 1997. From top to bottom, the modulus of the magnetic field $|\mathbf{B}|$ (nT); the three components of $\mathbf{B}$ in the GSE coordinates (nT); the three components of $\mathbf{B}$ in the LMN coordinates (nT); the modulus of the proton velocity $|\mathbf{V}_p|$ (km/s); the three components of $\mathbf{V}_p$ in the GSE coordinates (km/s), where $V_x$ has been shifted by +350 km/s; the proton number density $n_p$ (cm$^{-3}$); the proton temperature $T_p$ (eV); and the proton beta $\beta_p$. The exhaust interval is bounded by the two vertical dashed lines, which define $\mathbf{B}_1$ and $\mathbf{B}_2$, respectively.}
  \label{fig2}
\end{figure}

\begin{figure}
  \begin{center}
    \includegraphics[width=0.9\textwidth]{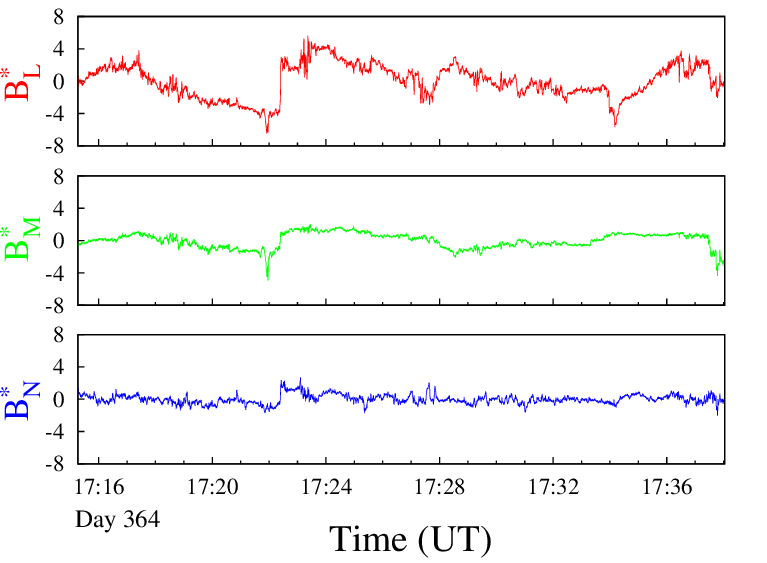}
  \end{center}
  \caption{The time series of $B_L^*$, $B_M^*$, and $B_N^*$, obtained by removing the trend computed by applying a third-order polynomial fit to the original data (event 1).}
  \label{fig3}
\end{figure}

\begin{figure}
  \includegraphics[height=0.9\textheight]{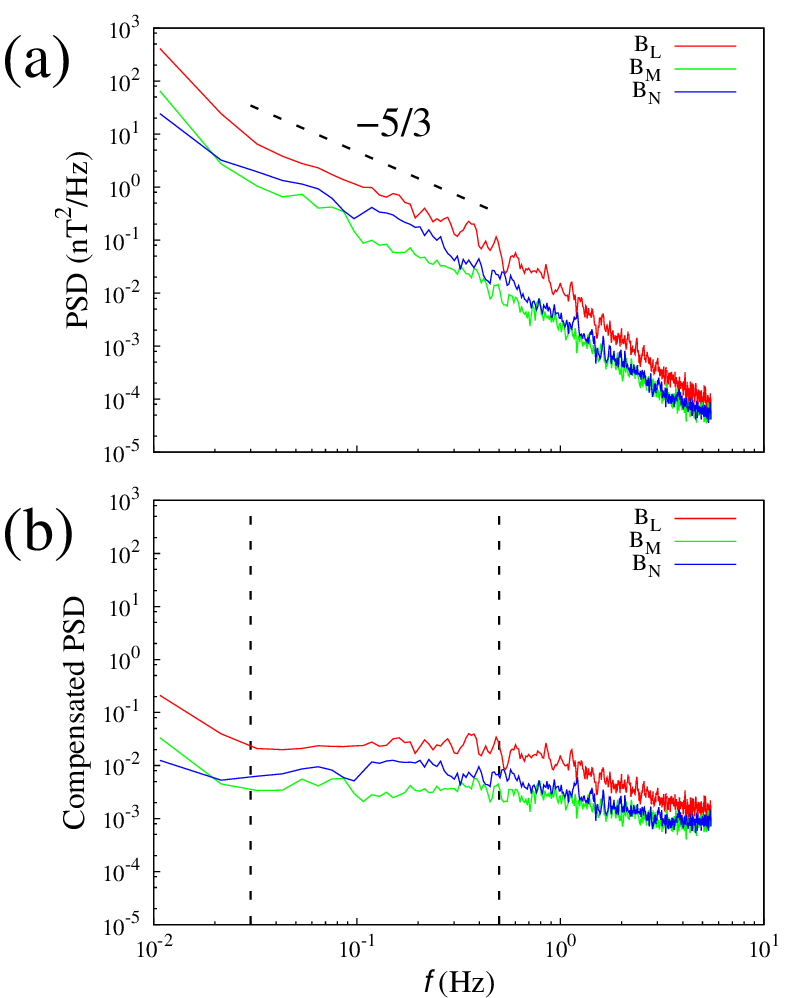}
  \caption{(a) Power spectral density (PSD) of the $B_L$, $B_M$, and $B_N$ components of the magnetic field (event 1). (b) Compensated PSD. The dashed vertical lines indicate the inertial subrange in which the compensated PSD is nearly horizontal.}
  \label{fig4}
\end{figure}

\begin{figure}
  \begin{center}
    \includegraphics[height=0.9\textheight]{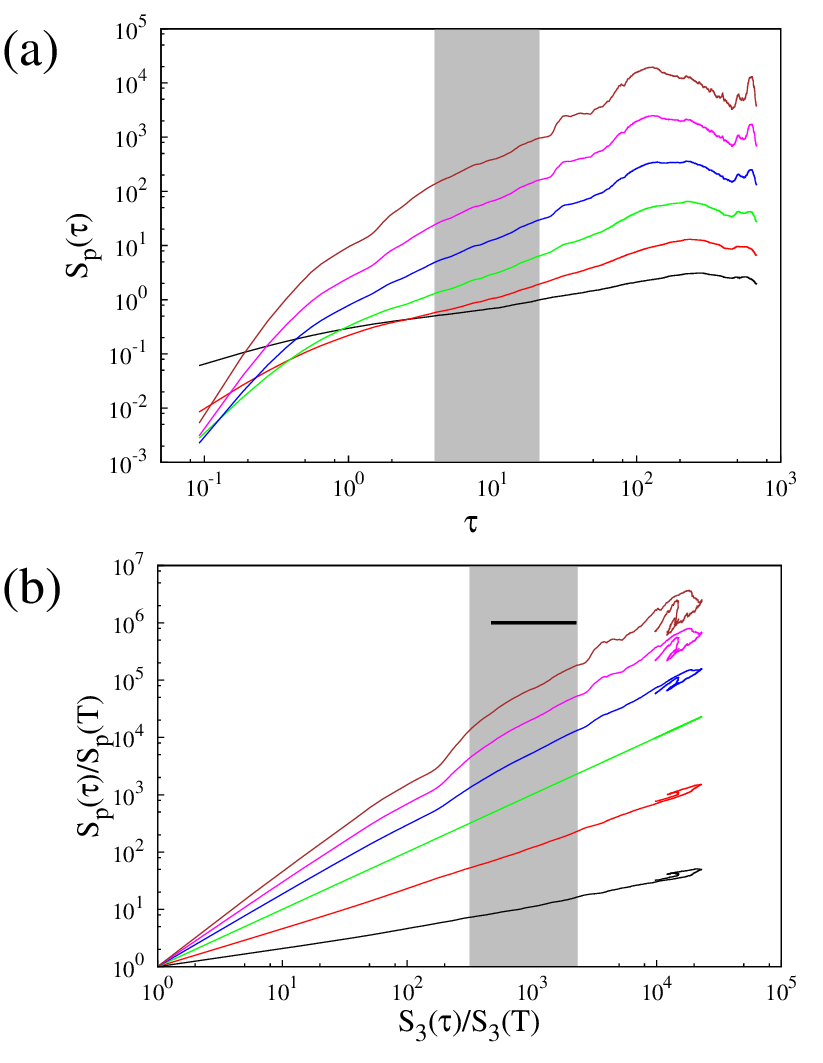}
  \end{center}
  \caption{(a) Structure functions as a function of $\tau$ computed from the $B_L$ component of the magnetic field (event 1), for $p = 1$ (black), $p = 2$ (red), $p = 3$ (green), $p = 4$ (blue), $p = 5$ (violet), and $p = 6$ (brown). The grey area represents the inertial subrange ($\tau \in [4, 21]$ s). (b) The structure functions after applying the Extended Self-Similarity technique. The horizontal line represents the original inertial subrange, and the grey area represents the extended scaling range ($S_3(\tau)/S_3(T) \in [320, 2302]$). The structure functions have been normalized to $S_p(T = 0.091 \sim$ 1/11 s). Note that the spectral break in Fig. \ref{fig4} occurs at scale $\tau \sim 2$ s, which corresponds to $S_3(\tau)/S_3(T)$ = 238 in panel (b).}
  \label{fig5}
\end{figure}

\begin{figure}
  \includegraphics[width=\textwidth]{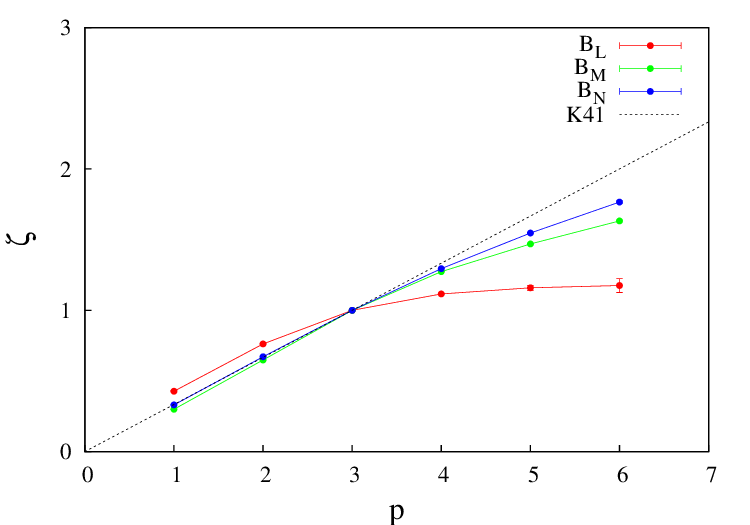}
  \caption{Scaling exponents for the $B_L$, $B_M$ and $B_N$ components at the exhaust region (event 1). The dotted line represents the K41 monofractal scaling with $\zeta(p)=p/3$. }
  \label{fig6}
\end{figure}

\begin{figure}
  \includegraphics[width=\textwidth]{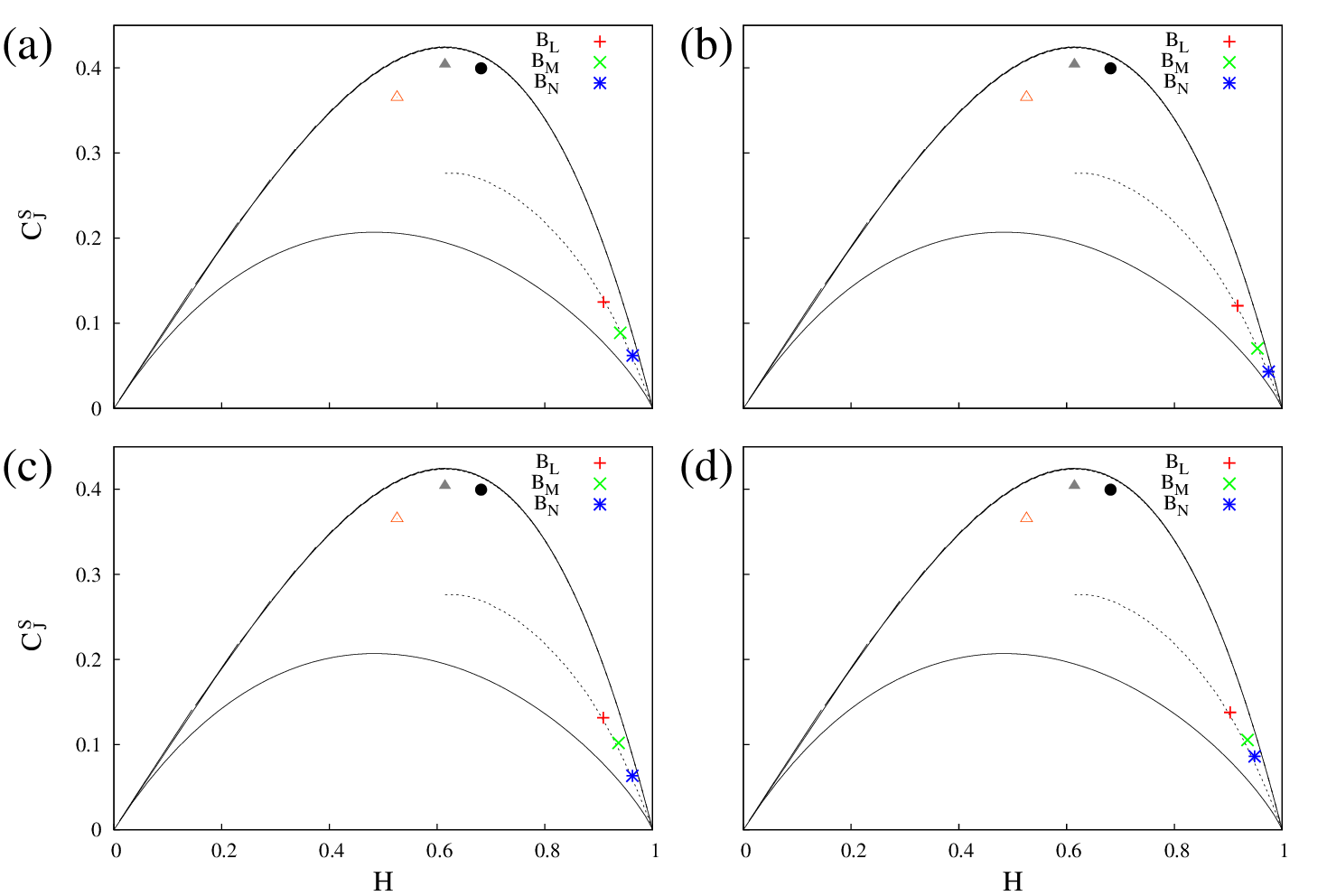}
  \caption{The $d$ = 5 Jensen-Shannon complexity plane for the $B_L$ (red plus symbol), the $B_M$ (green cross), and the $B_N$ (blue asterisk) components of the magnetic field during the reconnection exhaust interval of (a) event 1, (b) event 2, (c) event 3, and (d) event 4. The full black circle, open red triangle, and full grey triangle represent the chaotic time series of the logistic map, the skew tent map, and the H\'enon map, respectively (see the text for parameters). The crescent-shaped curves indicate the maximum and minimum values of $C_J^S$ for a given value of $H$, and the dotted line represents stochastic fractional Brownian motion.}
  \label{fig7}
\end{figure}

The exhaust intervals in Table \ref{table1} were carefully defined to avoid including the strong discontinuities due to the magnetic-field reversal near the boundaries of the reconnection exhaust (see the time series of $B_L$ in Fig. \ref{fig2}). We have also applied a detrending technique to further remove large-scale variations of the time series (see Fig. \ref{fig3}). Despite this, the higher degree of intermittency and complexity, and the lower degree of entropy, observed in the $B_L$ as compared with the other components, can still be due to the field reversal that occurs at a larger scale. Previous studies have pointed out evidence of a direct coupling between large-scale fluctuations and small-scale intermittency in the solar wind \citep{voros_etal:2006, miranda_etal:2018}. Therefore, the inertial-range coherent structures in the $B_L$ component within the reconnection exhaust can have their origin on the magnetic reconnection process that occurs at a larger scale. Note that this is in agreement with the main conclusion of \cite{chian_etal:2016}.

Our analysis of the three components of the magnetic field by the hybrid MVA suggests that intermittency and complexity-entropy vary with the field direction. Other characterizations of anisotropy in magnetic-field fluctuations in the solar wind, such as spectral anisotropy, have been demonstrated by several studies \citep[e.g.,][]{matthaeus_etal:1990, dasso_etal:2005, safrankova_etal:2021}. In that context, spectral anisotropy refers to the unequal distribution of energy between the wave-vectors directed parallel and perpendicular to the mean magnetic field. Energy spectra computed using single-spacecraft data displays unequal distribution of energy among magnetic-field components, which is termed variance anisotropy. However, this anisotropy observed using single-spacecraft data is not sufficient to demonstrate spectral anisotropy, and must be interpreted with caution. For example, calculated variances from single-spacecraft data can be misleadingly anisotropic even in the presence of an isotropic distribution of energy \citep{oughton_etal:2015}. A careful analysis of the variance anisotropy displayed by the energy spectra, and the different behavior of intermittency and complexity-entropy of magnetic-field components will be the focus of a future work.

In summary, in this paper, we analyzed the LMN components of the magnetic field at the exhaust of four reconnection events detected in the solar wind at 1 AU. The link between intermittency and complexity within the inertial subrange was demonstrated for the first event by computing the scaling exponents and the J-S index. For the four events, all components have $H$ and $C_J^S$ values within the stochastic region of the C-H plane. The $B_L$ component displays a higher degree of intermittency, lower entropy, and higher complexity than the $B_M$ and the $B_N$ components. Our results confirm that magnetic-field turbulence within reconnection exhausts are intermittent with various levels of multifractality in different directions, suggesting that nontrivial coherent structures are responsible for varying degrees of entropy and complexity. These results can contribute with additional constraints to the ongoing efforts on magnetic reconnection modelling and numerical simulation.

\acknowledgments
The authors are grateful to the reviewer for the valuable comments. R.A.M. acknowledges financial support from FAP DF (Brazil) under award number 180/2020, and DPI/DPG/UnB (Brazil). J.A.V. acknowledges funding by the National Agency for Research and Development (ANID - Chile) under FONDECYT award number 1190703, and by the Air Force Office of Scientific Research (US) under award number FA9550-20-1-0189. Numerical codes are freely available at {\tt https://gitlab.com/rmiracer}.

\bibliography{miranda_manuscript}

\clearpage

\begin{table*}
  \begin{center}
   \begin{tabular}{c|c|c|c}
      \hline
      Event & Interval start (UT) & Interval end (UT) & No. data points\\
      \hline
      1 & 30-12-1997 17:15:16 & 30-12-1997 17:38:02 & 14783\\
      2 & 14-11-1998 07:53:09 & 14-11-1998 08:12:10 & 11936\\
      3 & 23-11-1997 12:52:22 & 23-11-1997 13:07:28 & 8834\\
      4 & 02-02-2002 02:32:05 & 02-02-2002 02:34:34 & 3333\\
      \hline
    \end{tabular}
  \caption{Time interval and number of data points of four magnetic reconnection exhaust events.}
  \label{table1}
  \end{center}
\end{table*}

\clearpage

  \begin{sidewaystable*}
    \begin{center}
    \begin{tabular}{c|c|c|c|c|c|c}
      \hline
        & \multicolumn{2}{c|}{$d = 4$} & \multicolumn{2}{c|}{$d = 5$} & \multicolumn{2}{c}{$d = 6$}\\
      \cline{2-7}
         & H & $C_J^S$ & H & $C_J^S$ & H & $C_J^S$ \\
      \hline
      $B_L$ & 0.9361 $\pm 0.0002$ & 0.0725 $\pm$ 0.0002 & 0.9089 $\pm$ 0.0001 & 0.1250 $\pm$ 0.0002 & 0.8751 $\pm$ 0.0002 & 0.1951 $\pm$ 0.0004\\
      $B_M$ & 0.9599 $\pm$ 0.0005 & 0.0470 $\pm$ 0.0005 & 0.9400 $\pm$ 0.0005 & 0.0888 $\pm$ 0.0006 & 0.9114 $\pm$ 0.0007 & 0.1644 $\pm$ 0.0009\\
      $B_N$ & 0.9769 $\pm$ 0.0001 & 0.0292 $\pm$ 0.0002 & 0.9631 $\pm$ 0.0001 & 0.0621 $\pm$ 0.0002 & 0.9416 $\pm$ 0.0002 & 0.1178 $\pm$ 0.0008\\
      \hline
    \end{tabular}
    \end{center}
    \caption{The $(H, C_J^S)$ values of the $B_L$, $B_M$, and $B_N$ components of the magnetic field for $d$ = 4, 5, and 6, for event 1.}
    \label{table2}
  \end{sidewaystable*}

\clearpage

\clearpage

\clearpage

\clearpage

\clearpage

\clearpage

\clearpage

\end{document}